# Microstructural, vibrational, dielectric and ferroelectric properties correlation in hot-pressed PbMg$_{1/3}$Nb$_{2/3}$O$_3$ ceramics

**Adityanarayan H. Pandey** [a,b,*], **Surya M. Gupta** [a,b,*]

[a]*Homi Bhabha National Institute, Anushaktinagar, Mumbai-400094, India.*
[b]*Laser Materials Section, Raja Ramanna Centre for Advanced Technology, Indore-452013, India.*

**Vasant G. Sathe** [c], **Niranjan P. Lalla** [c]

[c]*UGC-DAE Consortium for Scientific Research, University Campus, Khandwa Road, Indore 452001, India.*

[*]*E.mail: anbp.phy@gmail.com* (Adityanarayan H. Pandey*); surya@rrcat.gov.in* (Surya *M. Gupta*)

**Abstract:** Microstrctural, vibrational, ferroelectric and dielectric properties investigation reveals non-stoichiometric chemical ordered regions (CORs) and relaxor like dielectric characteristics in lead magnesium niobate (PMN) ceramics uniaxially hot pressed (HP) at 800 °C to 1200 °C. Occurance of (½½½) superlattice reflections along <111> in <110> zone axis SAED pattern as well as the presence of bright nano-meter regions in corresponding dark-field images in 800 °C hot pressed PMN demonstrates formation of the CORs as soon as perovskite phase is formed during calcination and remains unaffected with HP sintering tempreature. Separation between two broad peaks of Nb-O-Nb stretching mode and the red-shift of O-B-O bending modes suggest an increase in the polar nano-size regions. Grain size dependence of the $\varepsilon_m$ and shifting of the $T_m$ from 278 K to 263 K is explained by the core-shell model. Polarization switching and impedance spectroscopy is observed to be consistent with the grain size analysis.

## 1. Introduction

Lead magnesium niobate (PMN) is first discovered in late 1950's by Smolenskii [1] and is still extensively investigated relaxor ferroelectric due to its unique structural, dielectric, polarization and microstructural properties. The PMN shows (i) frequency dependent diffuse dielectric maximum ($\varepsilon_m$ > 20000 at 1 kHz), which decrease with increasing frequency, (ii) shifting of temperature of the $\varepsilon_m$ ($T_m$) to lower side with decreasing frequency, (iii) inability to withstand a remnant polarization below the $T_m$, (iv) presence of polar nano-domains in the zero-field (ZF) condition, and (v) polarization-electric field (PE) hysteresis loss free loop at room temperature [2]. Current understanding about the PMN is immaculate on (i) presence of the paraelectric non-polar phase above the Burns temperature ($T_b$) ~ 600 K, which is demonstrated by deviation from linearity in temperature dependence of refractive index and by relating the $\varepsilon'(T)$ with the Curie-Weiss law [3] and (ii) coexistence of the polar nano-regions (PNRs) and the chemical order regions (CORs) below the $T_b$ [4]. Above the $T_b$, the Curie constant ~ $10^5$ has been determined, which is similar to that of normal ferroelectrics. Below the $T_b$, the PNRs are developed due to local structural distortion, which are permanently correlated dipole moments (giant dipoles) composed of many unit cells. Condensation of the soft transverse optical (TO) phonon mode is recently reported near the $T_b$ by elastic diffuse scattering of neutron [5]. Frequency dispersion in dielectric response of the PMN over wide range of temperature is believed to be a consequence of nano-size small ferroelectric polar clusters, i.e, the PNRs. Presently, it is believed that heterogeneities related to chemical, structural, and compositional are accountable for the unusual ferroelectric and dielectric properties of the PMN.

Several models are proposed to comprehend growth of the PNRs in the PMN [6]. Initially, the diffuse $\varepsilon_m$ peak is considered due to distribution of phase transitions ($T_c$) in crystal, but now it is recognized as response of the PNRs, which appears at the $T_b$ ~600 K. These dynamic PNRs grow in number as well as in size on cooling, which cause increase in the '$\varepsilon$' till it becomes static around T $\simeq$ 400 K. Finally, it is proposed that the PNRs ensemble into a nano domain ferroelectric state around 220 K (elliptical shape with maximum ~10 nm size) [7]. Phase analysis using the X-ray diffraction confirms Pm3m space group of cubic symmetry down to 5 K for the PMN [8]. Recently, by using synchrotron X-ray radiations, room temperature X-ray diffraction pattern along with Rietveld refinement has shown an improvement in the reliability fitting factor when coexistence of the Pm3m and R3m symmetry is considered for the PMN, which is also consistent with the general believe that structure of the PMN is the cubic globally and the rhombohedral locally [9].

There are two widespread models reported to explain origin of the PNRs [4,6,10,11]. According to the first model, i.e., Dipolar glass model, the PNRs are nano-size polar area, which is formed due to local phase transition and these PNRs are randomly distributed in a cubic background matrix. The second model assumes that the phase transition may be occurring similar to that of the normal ferroelectrics, but due to random electric field, the long range order breaks into smaller nano-domains. The quenched random electric field is originated due to the disorder present at the B-site cations, which is occupying equivalent crystallographic position in the perovskite lattice. Recently, frustrating dipolar interaction with the quenched disorder is suggested to obstruct the ferroelectric like transition in the PMN and ultimately leading to the PNRs [12].

Microstructural investigation by using transmission electron microscopy in the PMN has revealed coexistence of the CORs and the PNRs dispersed in the paraelectric host [4,7]. The CORs are believed to be source of the quenched random field, which destroy the long range polar ordering. Size of the CORs and separation between these regions is reported to 2-3 nm and ~ 2.5 nm, respectively [7,13]. These CORs are defined by "Random Layer model" where sucessive (111) planes are occupied by the $Nb^{5+}$ cations and mixture of the $Mg^{2+}$ and the $Nb^{5+}$ cations in a ratio of 2:1 giving rise to doubling of the unit cell size having Fm3m crystal structure [13,14]. Till now, the temperature at which the CORs forms in the PMN is not known but its presence has been observed up to 1000K using polarized Raman study [15]. Investigation of temperature dependent microstructural evolution [2,4,7] has revealed collapse of the elliptical shape ~10-20 nm PNRs around the CORs, which is also confirmed by diffused neutron scattering [5,16]. It seems, the interaction/coupling among the PNRs is hindered by the CORs resulting in restriction to achieve long range ordering [7]. Temperature dependent microstructure investigation has shown no effect on the number/size of the CORs in 15-800 K temperature range [7]. Thermal annealing is also reported to show no effect on the number/size of the CORs [16]. It is believed that the CORs and the PNRs are correlated, because applied external electric field results a ferroelectric rhombohedral structure in a poled PMN ceramic below 200 K, which is associated to the long range order [17].

In ferroelectric and relaxor ferroelectric ceramics, grain size has been reported to influence dielectric permitivity, spontaneous polarization and piezoelectric properties [18-23]. Reduction in real-part of permittivity ($\epsilon'$) with reduction in the grain size is generally explained by using a core-shell model, where presence of a dead grain boundary layer on the surface of each grain is assumed. The dead grain boundary layer in the PMN is reported to have lower permittivity and is probably composed of the pyrochlore phase or amorphous PbO. For the PMN, reduction in the grain size leads to decrease in the $\epsilon_m$ and shift of the $T_m$ to higher temperature is related to grain boundaries interference to the PNRs flipping. Carreaud *et al.* [21,22] reported the disappearance of the dielectric relaxation and vanishing of the correlations between the PNRs below 30 nm grain size of the PMN. It is also suggested that continuous reduction of the grain size should cause continuous disappearance of the correlation among the PNRs leading to non interacting PNRs state. Moreover, Grigalaitis *et al.*[23] has shown a progressive change from the Vogel-Fulcher like to the Arrhenius like (superparaelectric state) slowing down of the PNRs dynamics with reduction of the grain size from 2 μm to 15 nm. Interestingly, ferroelectric to relaxor transformation is also observed with reduction of the grain size in solid solution PMN-PT(65/35) [24]. Smaller correlation length is reported in small grains, which is responsible for transformation of long range to short range order [24].

Relaxor like dielectric characteristics of Ba $(Zr_x,Ti_{1-x})O_3$ system are recently reported, where polarizability difference between Zr and Ti ions has been stated crucial for the formation of the PNRs [25]. However, in the PMN, frustration interactions between the dipoles is expected for complex nano-domain structure. The nano scale heterogeneity is believed to be a fundamental part of relaxors, but the mechanisms of the CORs appearance and incapability of the PNRs growth below the $T_m$ in zero-field condition is still not clear. It is also not clear at what temperature the CORs develop in PMN and how the size of the CORs relate with the sintering temperature. The statistical distributions of the PNRs and the CORs is believed to play critical role in exhibiting relaxor behaviour of the PMN. The purpose of this study is to investigate origin of the CORs in hot pressed PMN and correlate hot-pressed sintering temperature with the size of the CORs and the relaxor properties. Present study clearly reveals the presence of the CORs and relaxor-like dielectric characteristics in the PMN ceramic hot pressed at 800 °C under 100 MPa, which is then comprehensively characterized for microstructural, dielectric and ferroelectric properties.

## 2. Experimental Details

The Columbite method is used for synthesising calcined powder of the PMN and details are described in ref [26]. Single perovskite phase is confirmed in the powder, which is calcined at 800 °C for 2 hours. The calcined powder is mixed with polyvinyl alcohol (PVA-3mol%) binder and pressed into cylinders (15 mm diameter with 5-6 mm thickness) using a uni-axial pressure of 200 MPa. These green ceramics are first heated at 450 °C for 4 hours to remove binder and then hot pressed (HP) under 100 MPa uniaxial pressure at different temperatures from 800 °C to 1200 °C for 2 hours. The PMN ceramic hot pressed at 800, 900, 1000, 1100 and 1200 °C are designated as HP800, HP900, HP1000, HP1100 and HP1200. The density of sintered pellets is determined by liquid displacement method (the Archimedes Method). The density of the hot-pressed ceramic is found to increase from 76% to 99% of the theoretical value with increasing HP sintering temperature. Room temperature X-ray diffraction (XRD) pattern of the polished ceramic is measured on a Rigaku diffractometer (Cu $K_α$ source with λ = 1.54 Å) using a scan rate of 2 °/min with 0.01° step size in the 2θ range 20-60$^o$. Crystal structure and the lattice parameters are calculated by Lebail fitting of the XRD pattern of the hot pressed ceramics using commercially available Fullprof software [27]. Grain size and its morphology is noted using field emission Scanning Electron Microscope (FE-SEM, Carl Zeiss, SIGMA) equipped with energy dispersive spectroscopy (Oxford Inca X-Act LN2 free) from a fractured surface. The fractured ceramic surface is sputtered with gold to obtain a thin layer before examining under the electron microscope. Average grain size is measured by the linear intercept method using 'imageJ' software. Transmission Electron Microscope (TEM) samples are constructed by ultrasonically drilling 3-mm discs, which have been mechanically polished to 100 μm. The centre portions of these discs are then further grounded to 10 μm by a dimpler, and then argon ion-milled (operating at 2-4 keV) to perforation. The TEM studies have been performed on a Phillips CM200 microscope operating at an accelerating voltage of 200 kV. Raman spectra of the hot pressed sintered ceramic is measured using LABRAM HR-800 spectrometer equipped with a 488 nm excitation source and a CCD detector, which gives real spectral resolution of better than 0.5 cm$^{-1}$. The deconvolution of the overlapping modes in the Raman spectra is carried out using Jandel 'peakfit' software. The fitting of the whole spectra is accomplished by using a Pseudo Voigt peak-shape functions (PV = $p$ x L + (1-$p$) x G, 0 ≤ $p$ ≤ 1, where L and G stand for Lorentzian and Gaussian, respectively) to deduce the characteristic parameters of all the Raman bands, e.g., intensity, full width at half maximum (FWHM) and peak position. For dielectric and PE loop measurements, the sintered disks are polished on different grades of emery papers to acquire parallel surfaces. The polished surfaces are ultrasonically cleaned to take off dust particles. The sample is sputter coated with gold followed by a thin coating of silver paste (dried at 450 $^o$C for 2 minutes) to assure good electrical contact. The dielectric property is measured using a Hewlett-Packard 4194A impedance analyser, which can cover a frequency range from 0.1 kHz to 100 kHz. For low temperature study, the sample is kept in a Delta Design 9023 test chamber, which can function between 90 K and 450 K. The temperature of ceramic sample is measured using a Eurotherm temperature controller via a K-type thermocouple fixed directly on the ground electrode of the sample fixture. The analyser, test chamber and Eurotherm temperature controller is interfaced with computer to collect data at 20 different frequencies while cooling at a rate of 2 $^o$C/min. For high temperature impedance spectroscopy measurement, sample is heated at 750 K in Lab-made set up and data is recorded by 6505B precision impedance analyzer (Wayne-kerr instrument, which can cover a frequency range of 20 Hz - 5 MHz) at selected temperatures while cooling.

Room temperature field induced polarization (PE hysteresis loop) is measured at 50 Hz using a Precision workstation of Radiant Technology, USA. For low temperature PE loop measurement, ceramic sample is kept in lab-made "cold finger" set up merged in liquid nitrogen and temperature of the sample is measured using a RTD sensor mounted directly on the ground electrode of the sample fixture. All the HP-sintered ceramics are tested up to maximum 15 kV/cm external field applied at 50 Hz. Application of more than 15kV/cm at lower frequencies are not carried out due to sample fixture limitation.

## 3. Results and Discussion

### 3.1. SEM

Microstructure images are recorded from the fractured surface of HP800 to HP1200 samples using scanning electron microscope and are shown in Fig. 1. For HP800, spherical-like grains and large open pores are visible, which is consistent with its density reported in Table 1. Increase in the HP sintering temperature reveals change in morphology of the grains from smooth grain surface to well defined facets of the grains having sharp grain boundaries. With increasing HP sintering temperature, the extent of open porosity reduces drastically above 1000°C. Figure 1(e) reveals well defined large grains with minimal open porosity for HP1200. Average grain size has been measured using the linear intercept method. Increase in the HP sintering temperature from 800 to 1200°C causes an increase in the average grain size from 1.1 μm to 2.8 μm.

### 3.2. XRD

Figure 2 compares room temperature X-ray diffraction (XRD) pattern of HP800 to HP1200 ceramics. All major peaks are indexed with JCPDS 27-1199 (perovskite phase of the PMN) confirming pseudo-cubic crystal structure having Pm3m space group [28]. A small amount of pyrochlore phase (JCPDS 37-0071, marked by *) has been observed for the HP800 and HP900 ceramics. The relative intensities of the (110) perovskite PMN peak ($I_{PMN}$) and the (222) pyrochlore peak ($I_{Pyro}$) have been used to determine volume fraction of the perovskite and the pyrochlore phase using Eq. 1, which are also presented in Table 1.

$$\text{Pyro/PMN (\%)} = \frac{(I_{Pyro/PMN})}{I_{Pyro}+I_{PMN}} \times 100 \qquad (1)$$

The pyrochlore phase < 2 wt.% for the HP800 and HP900 ceramics may be due to poor control of environment for lead oxide loss during HP sintering. Single perovskite phase is obtained in all the other PMN ceramics hot pressed above 900 °C, which is also evident in Fig. 2. From the structural analysis, lattice constant (a ~4.0467 Å) is determined considering Pm3m symmetry for the perovskite phase. No change in the lattice parameter is found with increasing HP sintering temperature, which is also consistent with the earlier reports [21].

The Full Width at Half Maximum (FWHM) of all the major peaks are found increasing with increase in the HP sintering temperature. Inset of the Fig. 2 shows variation of the FWHM for (111) and (211) peaks after subtracting the FWHM of standard (Si) sample (for removing instrumental broadening). The FWHM is observed to increase initially and then tends to saturate with increasing HP sintering temperature.

*3.3. TEM*

Presence of the CORs and the PNRs in HP sintered PMN ceramic is directly visualized using ⟨110⟩ zone selected area electron diffraction (SAED) pattern along with bright and dark field transmission electron microscope (TEM) images [29-32]. The TEM observations have been used to study the effect of HP sintering temperature on the local chemical ordering of the B-site cations. Figure 3(a-i) represents comparison of the bright-field, SAED pattern and dark-field imaging of HP800, HP900, and HP1200 ceramics. Figure 3(a-c) shows local random contrasts indicating presence of the polar nano-domains at room temperature. The average size of the PNRs is measured to be less than 5 nm, which is consistent with the earlier reports describing its presence below the $T_b$ ~650 K. In contrast to HP800, it is also observed that number-density of nano-regions is increased with increasing HP sintering temperature; implying larger size for the PNRs. Increase of the number-density of the PNRs is also consistent with Raman spectroscopic analysis as reported in the following section. These PNRs are much smaller in size than a typical FE domain, which agrees well with the earlier TEM study on the PMN and other Pb-based mixed perovskite relaxors [29,32].

Figure 3(d-f) compares the ⟨110⟩ zone SAED patterns corresponding to HP800, HP900, and HP1200 ceramics, respectively. Along with the allowed intense reflections, originating from the basic perovskite structure, extra weak spots at (½½½) reciprocal positions along <111> (F-spots, super-lattice reflections) are also clearly evident, as marked by small arrow. These F-spots signify local chemical-ordering in these samples. Such ordering also causes doubling of the unit cell. Presence of these F-spots even for the HP800 sample clearly reveals that nonstoichiometric ordering is inherent to the PMN ceramic. It occurs during calcination itself and remains unaffected with the high sintering temperature.

Any variation in the F-spots intensity signifies change in the size of CORs. Figures 3(d-f) do not show any vivid variation in the intensities of the F-spots with HP sintering temperature. The intensity of the F-spot is identical to that of the atmospheric pressure sintered PMN sample (at 1200 $^o$C) reported earlier [30-32]. Invariance in the intensity of the (½½½) super-lattice reflections with HP sintering implies no change in the size of the CORs, which is also confirmed further with dark field images. Figure 3(g-i) compares dark-field image observed at room temperature using the (½½½) super-lattice reflections for HP800, HP900 and HP1200 ceramics, respectively. In these images, the white spots represent nonstoichiometric 1:1 Mg/Nb ordered CORs, which are randomly dispersed in disordered matrix with its domain size confined to less than 5 nm. The stability of the CORs against thermal annealing is believed to relate with its inability to undergo phase-separation into perovskite and pyrochlore.

The TEM studies thus show the presence of the F-spots in HP800 sample, which is hot pressed at the temperature of the calcination, revealing that the nonstoichiometric CORs is an inherent part of the PMN and remains unaffected even after sintering. There is no change in the intensity of the (½½½) superlattice reflection and the size of nano-regions in dark-field images with increase in the HP sintering temperature. The consequences of enhancement of the PNRs on dielectric properties are discussed in section 3.5.

*3.4. Raman Spectroscopy*

Raman spectroscopy is used to investigate the local structures of the PMN, where the global symmetry is different from the local symmetry. The Raman spectrum of all the hot pressed PMN ceramics is compared in Fig. 4(a), which reveals many overlapping broad Raman bands, indicating the existence of different local structure and large disorder [33-38]. No evidence of disappearance of the old modes or appearance of new modes is observed with increasing HP sintering temperature, confirming the same macroscopic phase structure. These spectra are adjusted by the Bose factor and normalized by the intensity of 788 cm$^{-1}$ '$A_{1g}$' mode of the PMN. In consistency with the earlier report, the Raman spectra is de-convoluted using the Pseudo-Voigt peak shape function [36]. Figure 4(b, c) shows representative Raman spectra of the HP1200 de-convoluted into 14 peaks with varying FWHM, wave number and intensity. The soft modes at ~50 cm$^{-1}$ are not apparent, due to the experimental limitation. Similarly, the Raman spectra are de-convoluted for all HP800 to HP1200 samples. Figure 5(a-f) shows Raman shift and the FWHM of various modes against HP-sintering temperature.

For conveyance, the Raman spectra are divided into three regions. In region I, two modes near 100 and 150 cm$^{-1}$, as indicated by dotted line $B_1$ and $B_2$ in Fig. 4(a), are red shifted (from 110.0 to 100.1 cm$^{-1}$ and 152.3 to 148.3 cm$^{-1}$) with increasing HP sintering temperature and is also shown in Fig. 5(a). The red shift in $B_1$ and $B_2$ modes suggests reduction in the force constant of Pb-BO$_6$ bond with increasing HP sintering temperature, since these modes are assigned to relative motion of Pb-ions against the rigid oxygen octahedral involving motion of B-atom. The intensity and FWHM of ~150 cm$^{-1}$ mode is found to increase from 0.12 to 0.20 and 56.9 to 64.2 cm$^{-1}$, respectively with increasing HP sintering temperature suggesting an enhancement in coupling between Pb and B-site cations and is shown in Fig. 5(d). This is consistent with the intensity and FWHM of ~150 cm$^{-1}$ mode correlation with the dynamical coupling between the B-site cations and off-centred Pb in PST [39].

In region II, the wide band from 200 cm$^{-1}$ to 400 cm$^{-1}$ consists of four Raman modes, as marked $C_1$ to $C_4$ in Fig. 4(a). The $C_1$ and $C_2$ mode are found red shifted (from 223.1 & 269.4 cm$^{-1}$ (HP800) to 216.0 & 264.4 cm$^{-1}$ (HP1200)). The $C_3$ mode does not shift but $C_4$ mode is blue shifted (from 350.8 to 358.2 cm$^{-1}$) with increasing HP sintering temperature and can be seen in Fig. 5(b). The FWHM of $C_1$, $C_2$ and $C_3$ mode is increased ~6-8 cm$^{-1}$ but for $C_4$ mode it is increased ~24 cm$^{-1}$, as shown in Fig. 5(e). The Raman shifts of these peaks are consistent with the earlier reports [35-38]. The intensity of $C_1$, $C_2$ and $C_3$ mode is decreased ~15% but for $C_4$ it is increased ~30% with increasing HP sintering temperature. The $C_2$ mode is observed due to the off-centred B-cations displacement from its cubic positions in the PNRs and its shift has been related with distortion of the centre-symmetric structure. Shifting of the $C_2$ mode to higher frequency with La-doping in PMN [36] has been related to the cation moving towards the centre, which means decrease in the polar charge separation leading to reduction in the size of PNRs. It may be inferred that shifting of $C_2$ mode to lower frequency with increasing HP sintering temperature should cause shifting of B-site cation away from the centre leading to an increase in the distortion, which should result in larger sized PNRs. Raman activity of the $C_3$ and $C_4$ modes is initiated by dynamical off-centred structural fluctuations because of electron-phonon coupling. The intensity ratio of $C_4/C_3$ mode ($\rho$) is reported to relate with the correlation length of coherent Pb-shifts along <111> directions when vibrations modes in PbSc$_{0.5}$Ta$_{0.5}$O$_3$ and PbSc$_{0.5}$Nb$_{0.5}$O$_3$ are compared [40]. The value of "$\rho$" has been found to increase from 0.55 to 0.77 exhibiting enlargement of the coherence length of off-centred Pb-shifts with increasing HP sintering temperature, implying reduction in the disorder at the B-site.

In region III, two broad peaks in frequency range of ~500-800 cm$^{-1}$ are assigned to the Nb-O-Nb stretching mode, as designated by $D_1$ and $D_2$ (Fig. 4(a)). The origin of the Nb-O-Nb stretching mode is ascribed to the Nb$^{5+}$ ion displacement from the centre in NbO$_6$ octahedral cage, which evolves below the $T_B$ and responsible for the formation of the PNRs [34,35]. The separation between $D_1$, $D_2$ broad peaks is observed increasing upon cooling, which has been linked to the size of the static PNRs [34,35]. With increase of HP sintering temperature from 800 to 1200 $^o$C, peak position of the $D_1$ mode doesn't show any variation but the $D_2$ mode is blue shifted from 573.2 cm$^{-1}$ to 583.2 cm$^{-1}$. Thus, the separation between $D_1$ and $D_2$ increases from 76.7 cm$^{-1}$ to 86.2 cm$^{-1}$, suggesting an increase in the PNRs size, which is consistent with earlier reports where Gd or La-substitution in PMN has resulted in decrease in the separation between these two modes because of the reduced PNRs size [36]. The intensity ratio ($I_{D2}/I_{D1}$) increases from 2.52 (HP800) to 3.54 (HP1200) is also consistent with increase in size of the PNRs observed from red-shift of the O-B-O bending ($C_2$) mode.

In region III, the highest intensity mode near 780 cm$^{-1}$, designated as $A_{1g}$ mode, assigned to the stretching mode of the $B_1$-O-$B_2$ bond (here Nb-O-Mg). After fitting Raman spectra of all HP ceramics, it is observed that the peak position, intensity, and FWHM of the $A_{1g}$ mode (~788 cm$^{-1}$) do not show any variation with HP sintering temperature, which implies no change in the size of the B-site cation ordered region, which is consistent with TEM analysis.

*3.5. Dielectric Spectroscopy*

Temperature and frequency dependent dielectric properties measurements are carried out to study the influence of HP sintering temperature on the relaxor dielectric behaviour. The real (ε') and imaginary (ε") parts of complex dielectric constant (ε*) as a function of temperature for HP800 to HP1200 ceramics at selected frequencies are shown in Fig. 6(a-e). The ε'(T) and ε"(T) for all HP ceramics show broad $ε_m$ with strong frequency dispersion near $T_m$, i.e. the $T_m$ shifts progressively towards lower temperature with decreasing frequency. This frequency dispersion clearly indicates that the relaxation process of the dynamic PNRs occurs at multiple time scale [2]. Typical relaxor like dielectric characteristics has been observed for the HP800 ceramic. It may be noted that the PMN powder is calcined at 800$^o$C and the same powder is hot pressed at 800$^o$C. The relaxor like dielectric characteristic of the HP800 suggests that the local chemical heterogeneity remains unaffected with the HP sintering temperature. Figure 7(a,b) compares the ε'(T) and ε"(T) plots at 1 kHz frequency for all HP ceramics. Inset of the Fig. 7(b) shows that with increase of the HP sintering temperature, the $\varepsilon_m$ increases from 8800 (HP800) to 25000 (HP1200) and the $T_m$ shift from 278 K to 263 K, which is in good agreement with the earlier reports [21-23]. The $\varepsilon'_m$ and $T_m$ at 1 kHz for all the HP ceramics are displayed in Table 1. Increase in the $\varepsilon'_m$ with HP-sintering temperature is due to increase in number and/or size of the PNRs and interaction between them, which is because of increase in the average grain size from 1.1 µm to 2.8 µm. In the later part of this section, this is explained by considering the core-shell model. Figure 8(a) compares effect of the HP sintering temperature on the $\varepsilon'(T)$ curves in reduced form $\varepsilon'/\varepsilon'_m$ vs $T/T_m$ at 1 kHz frequency. The broadness around the $T_m$ is reduced with increasing HP sintering temperature, which implies reduction in the degree of cation's site disorder at the B-site. Decrease in the value of the $T_m$ with increasing HP sintering temperature is consistent with enhanced correlation among the PNRs. Relaxor dielectric characteristics are mainly expressed by the correlation length among the PNRs and degree of diffuseness. Interacting PNRs should grow on cooling and enhance their correlation length, leading to a ferroelectric long range ordered (for large correlation length) or short range ordered nano-domains (for smaller correlation length). For ferroelectrics in paraelectric regions, the temperature dependence of $\varepsilon'(T)$ can be related with the Curie-Weiss law ($\varepsilon'(T,\omega) = C/(T - \theta_{CW})$; where C is the Curie constant and $\theta_{CW}$ is the Curie-Weiss temperature, are constants. The fitted parameters are presented in Table 2. For relaxor ferroelectric, deviation from the Curie-Weiss fitting is reported below the $T_b$ [41] and is shown in Fig. 8(b). Recently, modified form of the Curie-Weiss relation (Eq. 2) has been shown to fit the $\varepsilon'(T)$ above the $\varepsilon_m$ [42].

$$\frac{\varepsilon_A(\omega)}{\varepsilon'(T,\omega)} = 1 + \frac{(T-T_A(\omega))^2}{2\delta_A^2} \qquad (2)$$

where $\varepsilon_A(>\varepsilon_m)$, $T_A(<T_m)$ and $\delta_A$ (called degree of diffuseness) are adjustable parameters, essentially independent of frequency and effective in long temperature range above the $\varepsilon_m$. Figure 8(c) shows fitting of the $\varepsilon'(T)$, above the $\varepsilon_m$, to Eq. 2 for HP1100 ceramic and the fitting parameters are stated in Table 2 along with that for other HP ceramics. The $\delta_A$ is observed to decrease from ~55 to 42.5 with increasing HP sintering temperature. Higher value of the $\delta_A$ is reported to signify enhancement of the degree of diffuseness [42]. Hence increase in HP sintering temperature results in decrease in the degree of diffuseness, which is because of lesser charge/chemical disorder at the B-site.

Reduction in the dielectric permittivity with decreasing grain size has been explained by a core-shell model [19], which assumes the presence of a dead layer encapsulating each grain, and permittivity of the layer is considered much lower than the grain. The dielectric constant and porosity of the sample as a function of HP sintering temperature is presented in Table 3. It may be noticed that the sample PMN-HP800 has sintered density much lower as compared to all other HP samples having sintered density > 95% of the theoretical value. In order to compare the dielectric constant with the grain size, the dielectric maximum is corrected for porosity present in the sample using the following Rushman and Striven equation [43]:

$$K_{corr} = K_{obs} \times \frac{(2+V_2)}{2(1-V_2)} \quad (3)$$

where $K_{obs}$ is the experimental value at the $T_m$ and $V_2$ is the volume fraction of porosity present in the sample. Here the porosity is assumed to be not interconnected [44]. The corrected dielectric maximum value for each sample is reported in Table 3. The dielectric constant dependency on the grain size for PMN is explained according to a logarithmic mixing rule [45]:

$$\ln K_{obs} = V_{shell} \ln K_{shell} + V_{core} \ln K_{core} \quad (4)$$

where $K_{shell}$ and $V_{shell}$ and $K_{core}$ and $V_{core}$ are the dielectric constant and volume of the shell and core regions, respectively. Assuming the $K_{core}$ of 28000 (single crystal of PMN) and a $K_{shell}$ of 300, the grain boundary thickness is calculated and presented in Table 3. It may be noticed that the thickness of the shell decreases from 70 nm to 20 nm and the thickness of the core is increased monotonically. A ratio of the core to shell thickness clearly show an increasing trend from 15 to 139 with increasing HP sintering temperature or grain size, which is consistent with the increased dielectric constant value.

The $T_m$ is also observed to shift to lower temperature with increasing grain size. This shift can be explained if grain boundary influence on the thermal properties of the PNRS is considered. In small grains, regions near to grain boundary may have more probability of restricting thermal activity of the larger polar regions, which are associated with the lower frequencies. This means, the dielectric permittivity will have more contribution from the thermally active smaller volume polar regions that are associated with the higher frequency. This result in shifting of the $T_m$ to higher frequency regime, which is consistent with the increase in the $T_m$ observed with decrease in the HP sintering temperature. A decrease in the dissipation factor for smaller grain sized PMN-HP800 sample is also consistent with this mechanism.

Further, strong frequency dependent dielectric behaviour of all the HP ceramics is a consequence of statistical distribution of the PNRs over wide temperatures range. Fitting of frequency dispersion of the $T_m$ to different models can reveal insight of the interaction among the PNRs. Reasonableness of the fitting parameters with the Vogel-Fulcher relation (Eq. 5) means frustrated interaction among the PNRs leading to freezing of the polar cluster dynamics at finite temperature ($T_f$) [46], whereas reasonableness to the power law of critical dynamics (Eq. 6) reveals interaction among the PNRs resulting into a critical slowing down of the PNRs dynamics at temperature ($T_g$) [47].

$$f = f_o e^{\frac{-E_a}{k_B(T_m - T_f)}} \qquad (5)$$

$$f = f_o \left(\frac{T}{T_g} - 1\right)^{zv} \qquad (6)$$

where $f_o$ is attempt frequency ($\tau_o = \omega_o^{-1}$ is the microscopic time associated with flipping of fluctuating dipole entities), $E_a$ is activation energy, $k_B$ is the Boltzmann constant, $T_f$ is the Vogel-Fulcher temperature, where freezing of polar cluster dynamics take place, $zv$ is critical dynamic exponent and $T_g$ is glass transition temperature. Figure 8(d) compares fitting of the $T_m$ vs $f$ plot to Eq. 5 and 6 for all the HP ceramics. The corresponding fitted parameters and the goodness of fit for HP800 to HP1200 are reported in Table 4. The open symbols represent the experimental data points and solid line represents the fitted curve. Accurate value of the $T_m$ is determined by fitting the $\varepsilon'(T)$ curve for each frequency in a narrow temperature range around the $T_m$. The fitting parameter $T_f$ of the HP1200 is near non-ergodic to ergodic transition reported for PMN [48]. Decrease in the $T_f$ from 249 to 216K with increase in the HP sintering temperature is due to reduction in the B-site cation disorder. The fitting parameters $T_f$, $f_o$ and $E_a$ for the HP1200 are consistent with the earlier report [32]. The attempt frequency ($f_o$) is found to increase from $10^{12}$ Hz to $10^{14}$ Hz with increasing HP sintering temperature. An increase of the activation energy from 0.055 (HP800) to 0.105 (HP1200) can be understood if reduction in the B-site cation disorder is considered, which leads to enhancement in the correlation between the PNRs.

The fitting of $T_m$ vs $f$ plot with Eq. 6 for all the HP ceramics is also carried out. The fitting parameters $f_o$, $zv$ are presented in Table 4. The fitted parameters of HP800 are $f_o \sim 8.6 \pm 7.1 \times 10^{15}$ Hz, $T_g = 264.5 \pm 0.6$ K, and $zv = 10.0 \pm 0.4$. With increasing HP sintering temperature, the $f_o$ remains unchanged $\sim 10^{15}$ Hz. Recently in PMN, nonergodic ferroelectric cluster glass ground state (also known as "super-dipolar" glass) is reported to develop at a static glass temperature ($T_g = 238K$) from high temperature PNRs ensemble under random electrostatic interaction [49]. The fitting parameters $T_g$ and $zv$ of all other HP ceramics are in agreement with the mesoscopic size of the PNRs nearby the $T_g$, and are reported in Table 4.

The parameter $T_g$ is decreased from 264 to 241 K with increasing HP sintering temperature, which suggests critical slowing down of dynamics of the PNRs ensemble (cluster) below this temperature. It may by noticed that the $T_g$ of the HP1200 agrees well with the earlier report [49].

Relaxation behaviour is investigated using Cole-Cole dielectric relaxation model because the $T_g$ or $T_f$ is considered as the temperature below which the distribution of relaxation time diverges. Figure 9(a,b) shows ε'(f) and ε"(f) of HP900 at selected temperatures in between 240 K to 280 K, respectively. Dispersion in the ε'(f) is decreased with increasing temperature. Consistently, a broad peak is detected in the ε" (f), which shifts to lower frequency with decreasing temperature. A symmetric and narrow distribution of relaxation time is reported in PMN-PZN-PSN relaxor ferroelectrics only at high temperature ($\geq T_{m(100Hz)}$ +15°C) [50]. Similarly, frequency dependent complex dielectric constant is fitted to the Cole-Cole relation [51];

$$\varepsilon^*(\omega) = \varepsilon_\infty + \frac{\Delta\varepsilon}{(1+i\omega\tau_0)^{1-\alpha}} \qquad (7)$$

where $\Delta\varepsilon = \varepsilon_s - \varepsilon_\infty$, $\varepsilon_s$ and $\varepsilon_\infty$ are low and high frequency limits of the ε, $\tau_o$ is the mean relaxation time, and the parameter α (0 < α < 1) specifies the width of distribution of relaxation time. Solid lines in Fig. 9(a,b) denote the fitting to Eq. 7, which shows that the Cole-Cole relations give a reasonable approximation to frequency dependent complex experimental data. Temperature dependence of the Cole-Cole parameters, Δε(T), α(T), and $\tau_o$ (T) are compared in Fig. 9(c-e) for all the HP ceramics. Figure 9(c) shows increase in the Δε(T) with decrease in the temperature, which is consistent with less thermal disordering effect at low temperature. Assuming same value of the higher frequency $\varepsilon_\infty$, the Δε(T) depends only on the $\varepsilon_s$, which shows a peak at the $T_c$, below which the dynamics of the PNRs are sluggish and not responding easily to the external stimuli. A shifts of the $T_c$ to lower temperature with increasing HP sintering temperature suggests that the PNRs are either larger in size or co-operative interaction is increasing with increasing HP sintering temperature. For HP900, the parameter α is smaller (~0.78) at 275 K and then saturates to ~0.9 around 240 K on cooling, which shows broadening in the distribution of relaxation times. Similar trend is detected for all other HP ceramics and is shown in Fig. 9(d). The parameter α at $T_{m(100Hz)}$ +15 °C is observed to decrease from 0.85 to 0.53 with increasing HP sintering temperature, which infers decrease in the width of distribution of relaxation time, increase in the size of the PNRs and its co-operative interaction. The mean relaxation time, $\tau_o$ increases from ~$10^{-7}$ s for HP800 to $10^{-3}$ s for HP1200 at 240 °C and remain nearly same $\tau_o$ ~$10^{-10}$ s at high temperature

~280 K, as shown in Fig. 9(e). This is consistent with an increase of the size of the PNRs, which agrees well with microstructural and Raman studies.

*3.6. Impedance spectroscopy*

To get insight of the dead layer around the grain, impedance spectroscopy has been carried out. Figure 10(a) shows a typical Nyquist plot of HP1100 ceramic at 700 K and its de-convoluted plots having two microstructural elements (grain and grain boundary) along with the fitting curve. It may be noticed that both the semi-circles are depressed, i.e., centre of the semi-circular arc is below the abscissa axis. Thus series combination of two parallel RQ circuit is considered instead of the RC circuit. It is well known for depressed semicircle, the capacitance is replaced with Q (also called constant phase element, CPE), which is the Johnscher capacitance, and is introduced to include contribution of mobile carrier's not only to conduction but also to polarization in a universal capacitor [52]. Universal dielectric response of a capacitor is defined as follows:

$$Q = C_o \omega^{n-1} \quad (8)$$

where $C_o$ is the capacitance at $\omega = 1$ rad/s, $\omega$ is the angular frequency and 'n' is exponent having value in between 0 and 1. The exponent "n" reflects deviation from an ideal capacitance. The "Q" is identical to a capacitance component when the exponent n is equal to one.

Equivalent circuit consisting of two RQ elements in series (as shown in the inset of the Fig. 10a for HP1100) is used to fit (solid line) the complex plots. Low and high frequency arcs are ascribed to the grain boundary and the grain of the HP ceramic, respectively. The resistances ($R_g$ and $R_{gb}$), capacitances ($C_g$ and $C_{gb}$) and exponent n ($n_g$ and $n_{gb}$) due to grain and grain boundaries are obtained from fitting of Nyquist plots with the aid of a commercial available software "Z-view" [53]. Radius of the de-convoluted arc in the complex impedance plot corresponds to DC-resistance of the micro-structural element [54]. Resistances corresponding to the grain ($R_g$) and grain boundary ($R_{gb}$) are evaluated at 700 K, 675 K and 650 K temperatures and compared in Fig. 10(b) for HP1100. The DC-resistance of the grain and grain boundary is found to increase with decreasing temperature for all the HP ceramics. Figure 10c compares the Nyquist plots for HP800 to HP1200 ceramics and its fitting with the proposed circuits and the calculated grain and grain boundaries resistance is compared in Table 3 for all HP-ceramics. With increasing HP sintering temperature, the grain and grain boundary resistance at 700 K is found to increase from ~$8.8 \times 10^4$ Ω to $1.8 \times 10^6$ Ω and from 4.5 x $10^4$ Ω to 3.7 x $10^6$ Ω, respectively. Similar trend is noticed at other two temperatures. No appreciable change in the capacitance is noticed with increase in the HP sintering temperature.

*3.7. Field Induced Polarization Switching*

Electric field induced polarization switching is recorded at 300 K, 220 K, and 180 K under a drive of 15 kV/cm at 50 Hz and is compared in Fig. 11(a-c) for all HP ceramics, respectively. The hysteresis loss free "s-shaped" loop with zero remnant polarization and tendency of saturation at large electric field is observed at 300 K for all HP PMN ceramics, displaying non-linear dependency of the polarization to the applied electric field. The maximum polarization ($P_{max}$) ~16.8 μC/cm$^2$ (at 15 kV/cm) for HP1200 ceramic is consistent with that of the reported air-sintered PMN ceramic [36], which is sintered at 1200 $^o$C for 2 hours and has average grain size ~7-8 μm. Upon decreasing HP sintering temperature from 1200 to 1000 $^o$C, small decrease in the $P_{max}$ from ~16.8 to ~16.3 μC/cm$^2$ has been noticed, but on further decrease in the HP sintering temperature to 800 $^o$C, the $P_{max}$ decreases drastically from 16.3 to ~9 μC/cm$^2$. This is attributed to pinning of the PNRs domain boundary by the grain boundaries of the small grains. The $P_{max}$ for all other HP ceramics are measured under similar conditions and is tabulated in table 1.

Typical ferroelectric like P-E hysteresis loop is observed at 180 K for all HP ceramics except HP800 and is shown in Fig. 11(d-f), which is consistent with development of the long range order when more than the threshold field (~2 kV/cm) is applied [20]. With decreasing HP sintering temperature from 1200 to 800 $^o$C, the $P_{max}$, remanent polarization ($P_r$) and coercive field ($E_c$) is decreased from ~30 to 9 μC/cm$^2$, 24 to 4 μC/cm$^2$ and 7.3 to 5.5 kV/cm, respectively. A drastic change in the $P_{max}$ and $P_r$ is observed with increasing HP sintering temperature from 800 to 900 $^o$C and after that the it increases gradually. It is important to note that HP1200 to HP900 samples are able to sustain the $P_r$ at 180 K, which suggests transformation from short range (nano domains) to long range (macroscale domains) ordering because the external applied electric field is more than the random field. The $P_r$ is found to increase with decreasing temperature until a peak is observed around 170 K and on further cooling, the $P_r$ and $P_{max}$ is decreased. Initial increase in the $P_r$, $P_{max}$ and $E_c$ with decreasing temperature is due to less thermal disordering effect on the interaction among the PNRs. With further decrease in thermal disorder, frustration interaction may be setting in near 170K leading to slow down of the PNRs dynamics, which causes reduction in the $P_r$ and $P_{max}$. For HP800, the lower $P_r$ strongly suggests incomplete alignment of the polar nano domains, which may be due to reduced co-operative interaction among the PNRs and restriction in flipping or modifying size of the PNRs near to the grain boundaries.

4. Conclusion

Grain size dependence on the $\varepsilon_m$ and $T_m$ has been analysed using the core-shell model, which reveals increase in the ratio of core to shell region thickness with increasing grain size. Raman spectroscopy investigation has revealed an increase in the local distortion, which is not detected in X-ray diffraction pattern. Typical relaxor like dielectric characteristics, Super-lattice reflection (½½½) in <110> zone SAED pattern and $A_{1g}$ mode corresponding to chemical ordered regions has been observed in HP-800 sample, which confirms formation of the CORs during the calcined stage. Frequency and temperature dependence of the dielectric constant reveal decrease in the distribution width of the relaxation time and the degree of diffuseness, which imply increase in size of the PNRs and their co-operative interaction with increasing HP sintering temperature.

Acknowledgements

Authors are grateful to Rashmi Singh for FESEM measurement, Prem Kumar for XRD measurement, Priti Mahajan for the help during TEM sample preparation, Ajay Rathore for the help during Raman measurement, Ashok Bhakar for impedance spectroscopy measurement and Gurvinderjit Singh for useful discussions. Mr. Pandey acknowledges Homi Bhabha National Institute, India for research fellowship.

**Tables and Figures Captions**

**Table 1.** Variation of the lattice parameter, % pyrochlore phase, density, grain size, dielectric properties, and field induced polarization of hot pressed PMN ceramics.

**Table 2.** Parameters of the Curie-Weiss law (Eq. 3) and Quadratic law (Eq. 4) for various HP sintered PMN ceramics.

**Table 3.** Parameters of the **(a)** Vogel-Fulcher (Eq. 5) and **(b)** critical slowing down model (Eq. 6) fit of frequency dependent $T_m$ for various HP sintered PMN ceramics.

**Table 4.** Corrected dielectric constant, grain and grain boundary thicknesses calculated using series and logarithmic mixing rules for HP PMN ceramics.

**Table 5.** Grain and grain boundary resistance ($R_g$ ($\Omega$) and $R_{gb}$ ($\Omega$)) of HP-PMN ceramics obtained from equivalent circuit fitting

**Fig. 1.** FESEM micrograph of fractured surface of hot pressed PMN ceramics sintered at different temperatures a) HP800, b) HP900, c) HP1000, d) HP1100, and e) HP1200.

**Fig. 2.** Comparison of XRD patterns of Hot pressed PMN ceramics sintered at different temperatures from 800 $^o$C to 1200 $^o$C (HP800 to HP1200) and indexed using JCPDS 27-1199; secondary pyrochlore phase present in HP800 and HP900 are highlighted using '*'. Inset shows FWHM of (111) and (211) peak (after subtracting instrumental broadening) against HP-temperature.

**Fig. 3. (a-i)** Comparion of the bright field images **(a-c)**, selected area electron diffraction along <110> unit axis **(d-f)** and dark field images **(g-i)** for HP800, HP900 and HP1200 ceramics. Presence of superlattice reflection along ½ <111> axis is shown by an arrow.

**Fig. 4. (a)** Comparison of room temperature Raman spectra of various HP PMN ceramics, **(b, c)** Deconvolution of Raman spectra of HP1200; plus, marks represent the raw data, solid red line is the fitted data using Pseudo-Voigt peaks (green color) and the linear background line is given for clarity.

**Fig. 5. (a-c)** HP sintering temperature dependence of few frequency modes, and **(d-f)** HP sintering temperature dependence of FWHM of various modes for PMN ceramics.

**Fig. 6.** Real and imaginary parts of complex dielectric constant as function of temperature at selected frequencies of hot pressed PMN ceramics sintered at different temperatures, **(a)** HP800, **(b)** HP900, **(c)** HP1000, **(d)** HP1100, and **(e)** HP1200.

**Fig. 7.** Comparison of temperature dependence of **a)** dielectric constant and **b)** loss tangent at 1 kHz frequency; inset plots $\varepsilon_m$ and $T_m$ at 1 kHz frequency as a function of hot pressed sintering temperature. Inset of the Fig. 7b compares the $\varepsilon_m$ and $T_m$ with the HP sintering temperature.

**Fig. 8.** **(a)** Normalized curve ($\varepsilon/\varepsilon_m$ vs $T/T_m$) of hot pressed PMN ceramics sintered from 800 °C to 1200 °C, **(b)** Representative fitting ($1/\varepsilon$ vs T) of HP1100 ceramic sample by Curie-Weiss Law, **(c)** Representative fitting ($1/\varepsilon$ vs T) of HP1100 ceramic sample by Eq. 2; inset of Fig. 8(c) shows $\delta_A$ variation of HP sintering temperature, **(d)** Comparison of the Vogel − Fulcher (red curve), and critical slowing down model (blue curve) for all HP PMN ceramics where open symbols represent the temperature of $\varepsilon_m(T_m)$ in the frequency range 0.1 – 100 kHz.

**Fig. 9.** **(a, b)** Frequency dependence of real ($\varepsilon'$) and imaginary ($\varepsilon''$) parts of complex dielectric constant at selected temperatures between 240 K - 280 K for HP 900 PMN ceramic sample, where symbols represent the experimental data points and red solid lines are fitting to Eq. (5), **(c-e)** Variation of the Cole-Cole fitting parameter: **(c)** $\Delta\varepsilon$, **(d)** $\alpha$, and **(e)** $\tau_o$ with temperature for HP PMN ceramics sintered from 800 °C to 1200 °C.

**Fig. 10.** **(a)** Impedance spectra of HP1100 measured at 700 K (black solid spheres) is fitted using equivalent circuit scheme using grain and grain boundary contributions, **(b)** Temperature evolution of impedance spectra of HP1100 measured at 650 K, 675 K, and 700 K fitted using equivalent circuit scheme, **(c)** Impedance spectra of different HP sintered ceramics measured at 700 K fitted using equivalent circuit scheme

**Fig. 11.** **(a-c)** Comparison of PE hysteresis loop of HP sintered PMN ceramics (from 800 °C to 1200 °C) at E = 15 kV/cm and *f* = 50 Hz **(a)** T = 300, **(b)** T = 220 K, and **(c)** T = 180 K, **(d-f)** Comparison of temperature dependent parameters **(d)** $P_{max}$, **(e)** $P_r$, and **(f)** $E_c$ of HP PMN ceramics.

**Table 1.** Variation of the lattice parameter, % pyrochlore phase, density, grain size, dielectric properties, and field induced polarization of hot pressed PMN ceramics.

| PMN Samples | Lattice Parameter (Å) | % Pyro | Density % of Theoretical | Grain Size (µm) | $f \sim 1$ kHz | | Dissipation factor @ $T_m$ | $P_{max}$ (µC/cm$^2$) E = 15 kV/cm |
|---|---|---|---|---|---|---|---|---|
| | | | | | $\varepsilon'_m$ | $T_m$ (K) | | |
| HP800 | 4.048 | 1.7 | 78% | 1.1 | 8800 | 278 | 0.0195 | 9.0 |
| HP900 | 4.047 | 1.9 | 95% | 1.2 | 19900 | 271 | 0.0264 | 13.6 |
| HP1000 | 4.047 | - | 98% | 1.3 | 21000 | 268 | 0.0350 | 16.3 |
| HP1100 | 4.047 | - | 99% | 1.5 | 24500 | 266 | 0.0361 | 16.6 |
| HP1200 | 4.047 | - | 99% | 2.8 | 25000 | 263 | 0.0445 | 16.8 |

**Table 2.** Parameters of the Curie-Weiss law and Quadratic law (Eq. 2) for various HP sintered PMN ceramics.

| PMN Samples | Curie-Weiss Law | | Quadratic Law | | |
|---|---|---|---|---|---|
| | $C \times 10^5$ (K$^{-1}$) | $\theta_{CW}$ (K) | $T_A$ (K) | $\varepsilon_A$ | $\delta_A$ |
| HP800 | 1.1719 | 380.3 | 268.7 | 8990 | 57.7 |
| HP900 | 2.0153 | 387.5 | 259.4 | 20773 | 50.5 |
| HP1000 | 1.7204 | 381.6 | 255.5 | 22061 | 48.9 |
| HP1100 | 1.8901 | 383.8 | 246.9 | 26739 | 47.5 |
| HP1200 | 1.5455 | 387.4 | 246.7 | 26723 | 44.1 |

**Table 3.** Corrected dielectric constant, grain and grain boundary thicknesses calculated using logarithmic mixing rules for HP PMN ceramics.

| Sample | $K_{obs}$ (1kHz) | Porosity (%) | $K_{corr}$ (1kHz) | Average Grain size (μm) | Th-Core (μm) | Th-Shell (nm) | Ratio core/shell |
|---|---|---|---|---|---|---|---|
| HP-800 | 8800 | 22 | 12523 | 1.1 | 1.030 | 70 | 15 |
| HP-900 | 19900 | 5 | 21471 | 1.2 | 1.176 | 24 | 49 |
| HP-1000 | 21000 | 2 | 21642 | 1.3 | 1.275 | 25 | 51 |
| HP-1100 | 24500 | 1 | 24871 | 1.5 | 1.486 | 14 | 106 |
| HP-1200 | 25000 | 1 | 25378 | 2.8 | 2.780 | 20 | 139 |

**Table 4.** Parameters of the **(a)** Vogel-Fulcher (Eq. 6) and **(b)** critical slowing down model (Eq. 7) fit of frequency dependent $T_m$ for various HP sintered PMN ceramics.

| PMN Samples | Vogel-Fulcher model (Eq. 5) | | | Adj. R-square |
|---|---|---|---|---|
| | $T_f$ (K) | $E_a$ (eV) | $f_o$ (Hz) | |
| HP800 | 248.8(±1.6) | 0.055(±0.006) | 2.72(±1.08) x $10^{12}$ | 0.99956 |
| HP900 | 241.9(±0.9) | 0.049(±0.003) | 2.83(±1.58) x $10^{11}$ | 0.99982 |
| HP1000 | 230.6(±1.2) | 0.076(±0.005) | 1.02(±0.68) x $10^{13}$ | 0.99987 |
| HP1100 | 220.1(±1.3) | 0.097(±0.003) | 6.09(±3.9) x $10^{13}$ | 0.99991 |
| HP1200 | 216.4(±1.7) | 0.105(±0.007) | 2.26(±1.0) x $10^{14}$ | 0.99987 |

| PMN Samples | Critical slowing down model (Eq. 6) | | | Adj. R-square |
|---|---|---|---|---|
| | $zv$ | $T_g$ (K) | $f_o$ (Hz) | |
| HP800 | 10.0(±0.4) | 264.5(±0.6) | 8.56(±7.06) x $10^{15}$ | 0.99967 |
| HP900 | 8.9(±0.2) | 257.6(±0.3) | 2.96(±1.12) x $10^{14}$ | 0.99988 |
| HP1000 | 10.8(±0.4) | 250.7(±0.6) | 2.49(±1.32) x $10^{15}$ | 0.99984 |
| HP1100 | 11.7(±0.3) | 243.8(±0.6) | 2.55(±1.04) x $10^{15}$ | 0.99991 |
| HP1200 | 12.3(±0.4) | 241.0(±0.8) | 6.55(±3.44) x $10^{15}$ | 0.99987 |

**Table 5.** Grain and grain boundary resistance ($R_g$ ($\Omega$) and $R_{gb}$ ($\Omega$)) of HP-PMN ceramics obtained from equivalent circuit fitting.

| Ceramics | | 700 K | 675 K | 650 K |
|---|---|---|---|---|
| HP800 | $R_g$ ($\Omega$) | 8.81 x 10$^4$ | 1.50 x 10$^5$ | 3.69 x 10$^5$ |
| | $R_{gb}$ ($\Omega$) | 4.55 x 10$^4$ | 1.28 x 10$^5$ | 2.12 x 10$^5$ |
| HP900 | $R_g$ ($\Omega$) | 3.20 x 10$^5$ | 3.20 x 10$^5$ | 1.04 x 10$^6$ |
| | $R_{gb}$ ($\Omega$) | 8.2 x 10$^4$ | 2.77 x 10$^5$ | 1.48 x 10$^5$ |
| HP1000 | $R_g$ ($\Omega$) | 4.65 x 10$^5$ | 7.06 x 10$^5$ | 2.18 x 10$^6$ |
| | $R_{gb}$ ($\Omega$) | 2.14 x 10$^5$ | 7.84 x 10$^5$ | 1.00 x 10$^6$ |
| HP1100 | $R_g$ ($\Omega$) | 8.03 x10$^5$ | 6.21 x 10$^5$ | 2.93 x 10$^6$ |
| | $R_{gb}$ ($\Omega$) | 2.88 x 10$^5$ | 9.51 x 10$^5$ | 1.64 x 10$^5$ |
| HP1200 | $R_g$ ($\Omega$) | 1.84 x10$^6$ | 7.08 x 10$^6$ | 1.55 x 10$^7$ |
| | $R_{gb}$ ($\Omega$) | 3.69 x10$^6$ | 1.30 x 10$^6$ | 9.783 x 10$^5$ |

**Figure 1**

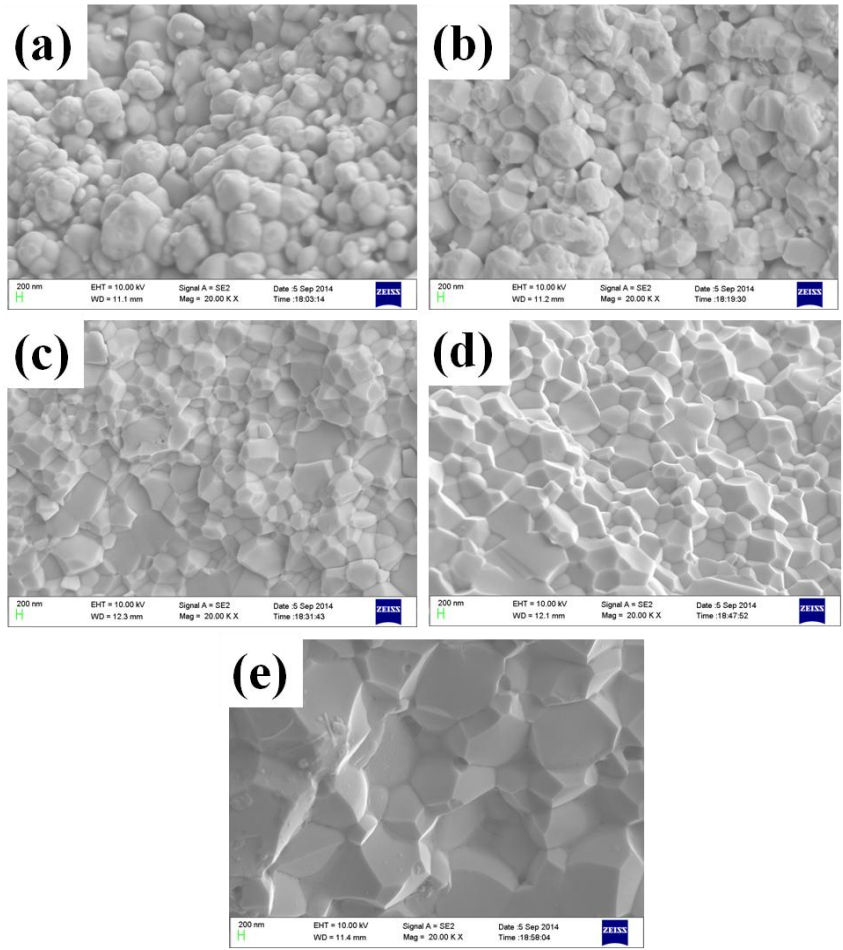

**Fig. 1.** FESEM micrograph of fractured surface of hot pressed PMN ceramics sintered at different temperatures a) HP800, b) HP900, c) HP1000, d) HP1100, and e) HP1200.

**Figure 2**

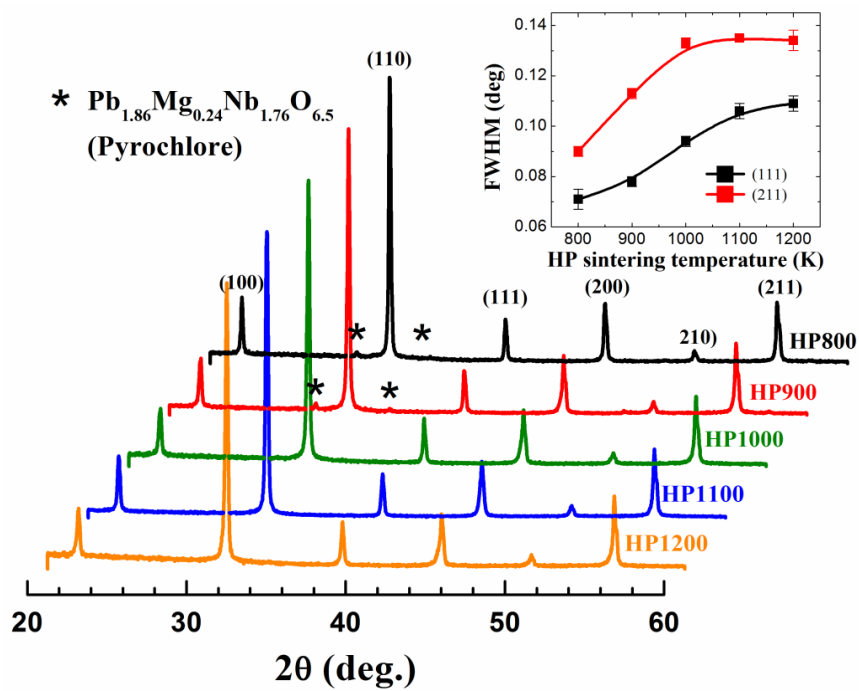

**Fig. 2.** Comparison of XRD patterns of Hot pressed PMN ceramics sintered at different temperatures from 800 °C to 1200 °C (HP800 to HP1200) and indexed using JCPDS 27-1199; secondary pyrochlore phase present in HP800 and HP900 are highlighted using '*'. Inset shows FWHM of (111) and (211) peak (after subtracting instrumental broadening) against HP-temperature.

**Figure 3**

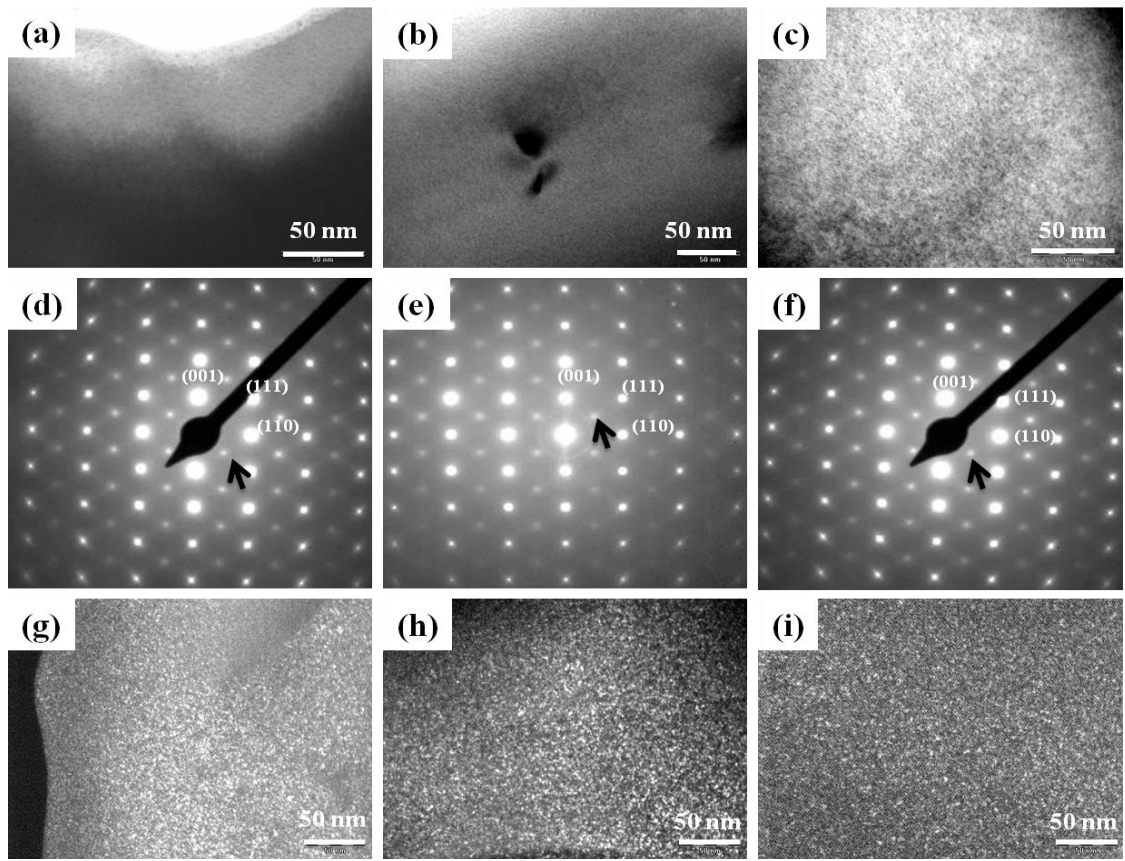

**Fig. 3. (a-i)** Comparion of the bright field images **(a-c)**, selected area electron diffraction along <110> unit axis **(d-f)** and dark field images **(g-i)** for HP800, HP900 and HP1200 ceramics. Presence of superlattice reflection along ½ <111> axis is shown by an small arrow.

**Figure 4**

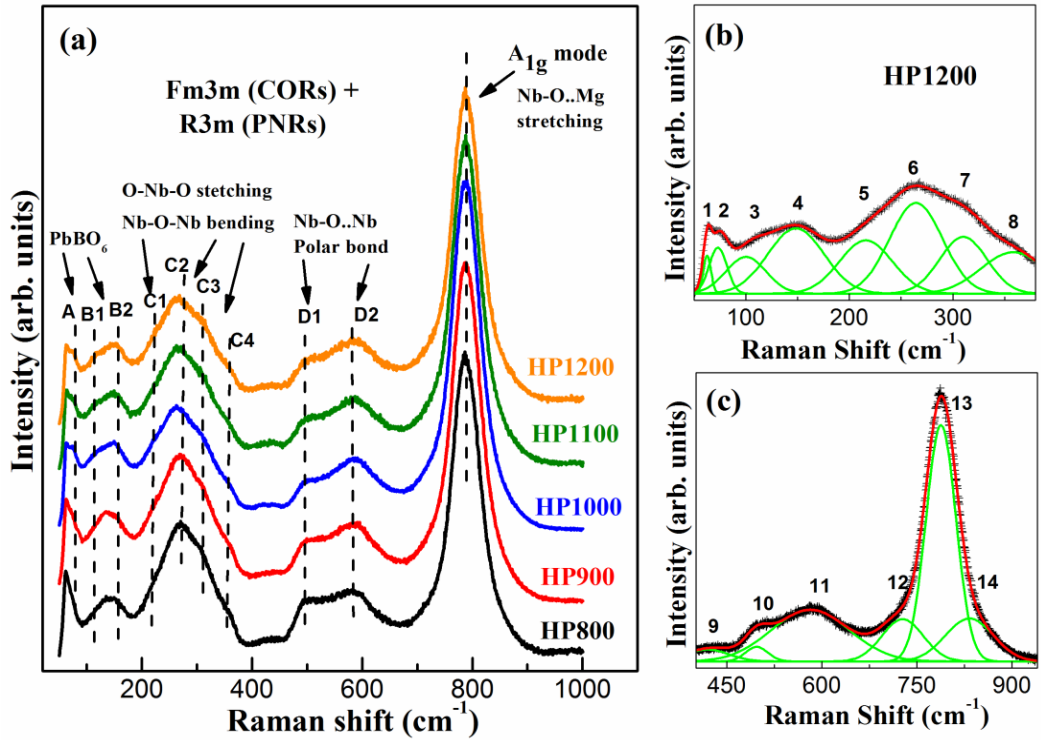

**Fig. 4. (a)** Comparison of room temperature Raman spectra of various HP PMN ceramics, **(b, c)** Deconvolution of Raman spectra of HP1200; plus, marks represent the raw data, solid red line is the fitted data using Pseudo-Voigt peaks (green color) and the linear background line is given for clarity.

**Figure 5**

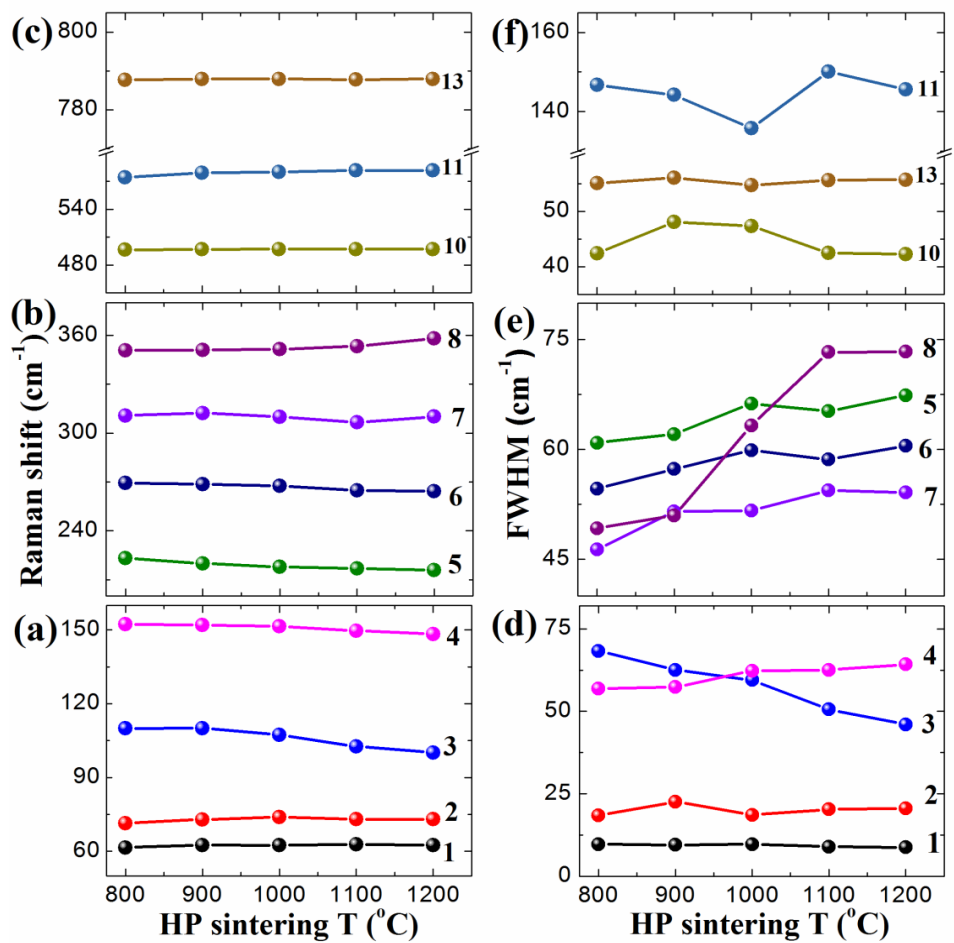

**Fig. 5. (a-c)** HP sintering temperature dependence of few frequency modes, and **(d-f)** HP sintering temperature dependence of FWHM of various modes for PMN ceramics.

**Figure 6**

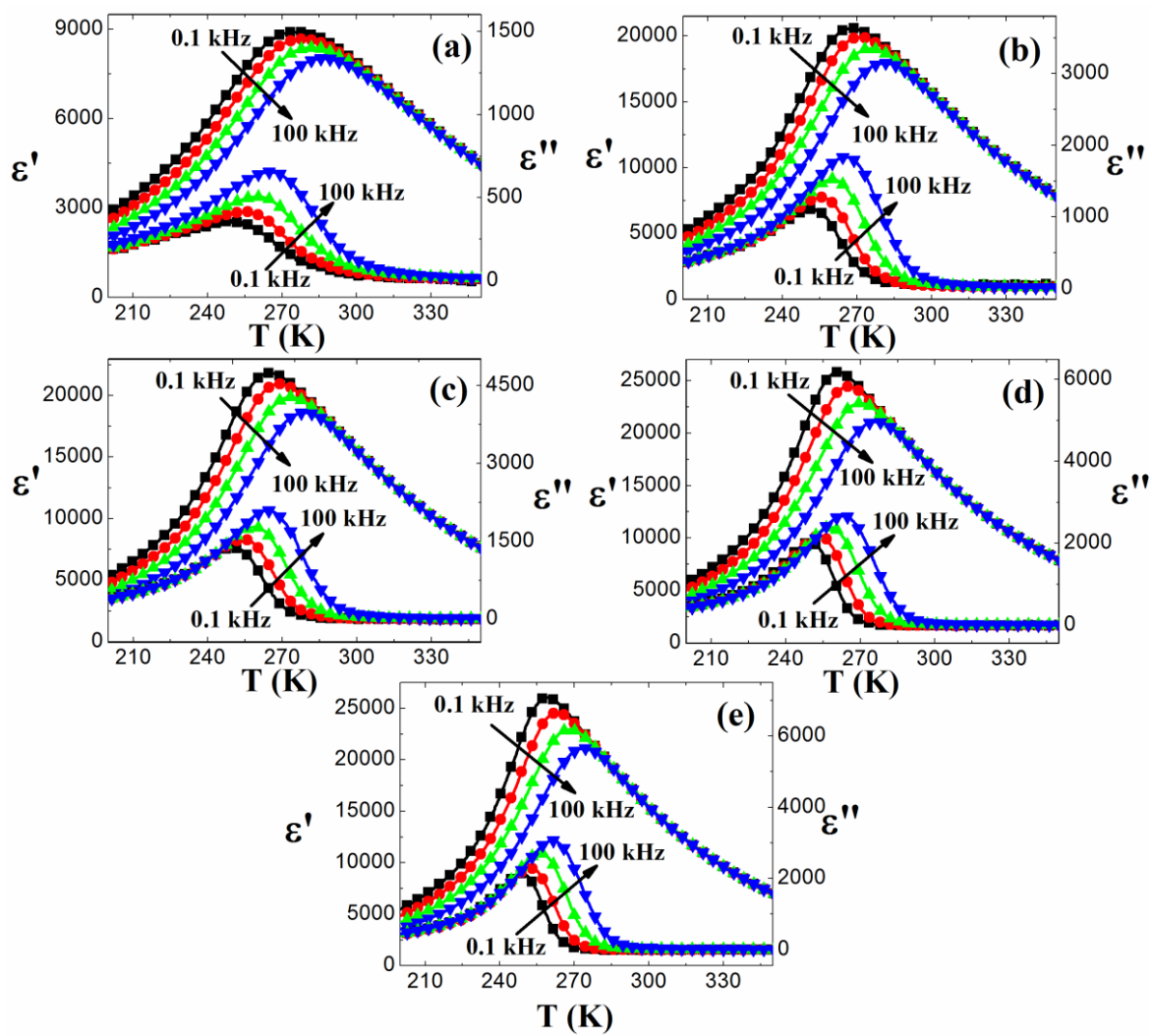

**Fig. 6.** Real and imaginary parts of complex dielectric constant as function of temperature at selected frequencies of hot pressed PMN ceramics sintered at different temperatures, **(a)** HP800, **(b)** HP900, **(c)** HP1000, **(d)** HP1100, and **(e)** HP1200.

**Figure 7**

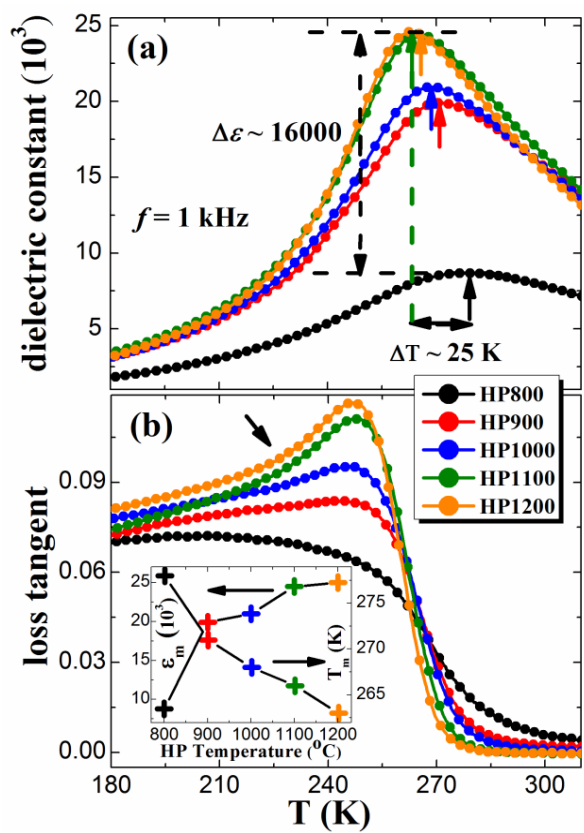

**Fig. 7.** Comparison of temperature dependence of **a)** dielectric constant and **b)** loss tangent at 1 kHz frequency; inset plots $\varepsilon_m$ and $T_m$ at 1 kHz frequency as a function of hot pressed sintering temperature. Inset of the Fig. 7b compares the $\varepsilon_m$ and $T_m$ with the HP sintering temperature.

**Figure 8**

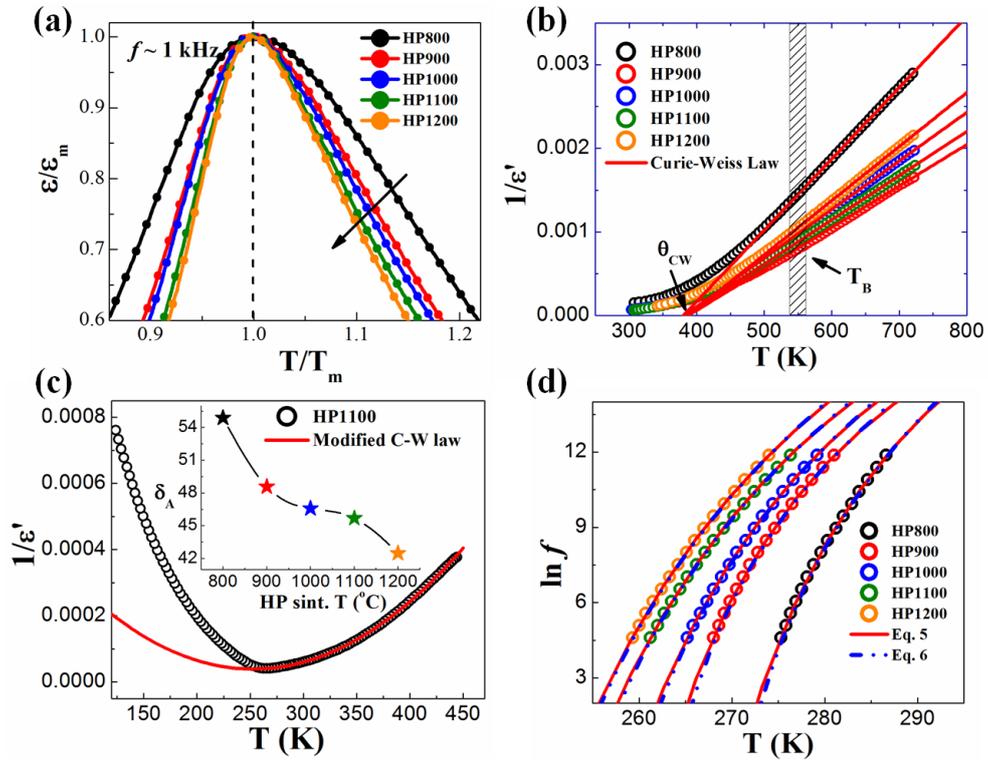

**Fig. 8.** **(a)** Normalized curve ($\varepsilon/\varepsilon_m$ vs $T/T_m$) of hot pressed PMN ceramics sintered from 800 °C to 1200 °C, **(b)** Representative fitting (1/ε vs T) of HP1100 ceramic sample by Curie-Weiss Law, **(c)** Representative fitting (1/ε vs T) of HP1100 ceramic sample by Eq. 2; inset of Fig. 8(c) shows $\delta_A$ variation of HP sintering temperature, **(d)** Comparison of the Vogel − Fulcher (red curve), and critical slowing down model (blue curve) for all HP PMN ceramics where open symbols represent the temperature of $\varepsilon_m(T_m)$ in the frequency range 0.1 – 100 kHz.

**Figure 9**

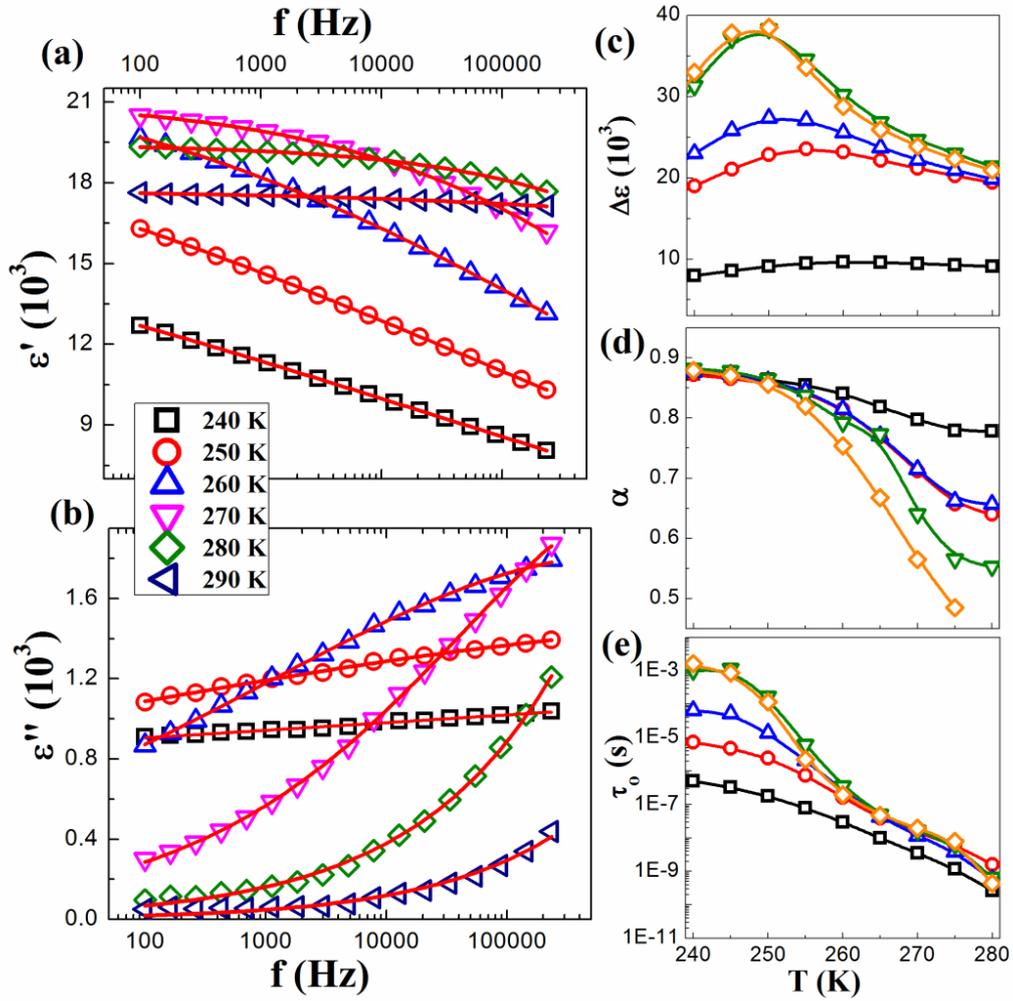

**Fig. 9.** (**a, b**) Frequency dependence of real (ε') and imaginary (ε") parts of complex dielectric constant at selected temperatures between 240 K - 280 K for HP 900 PMN ceramic sample, where symbols represent the experimental data points and red solid lines are fitting to Eq. (5), (**c-e**) Variation of the Cole-Cole fitting parameter: (c) Δε, (d) α, and (e) $\tau_o$ with temperature for HP PMN ceramics sintered from 800 $^o$C to 1200 $^o$C.

**Figure 10**

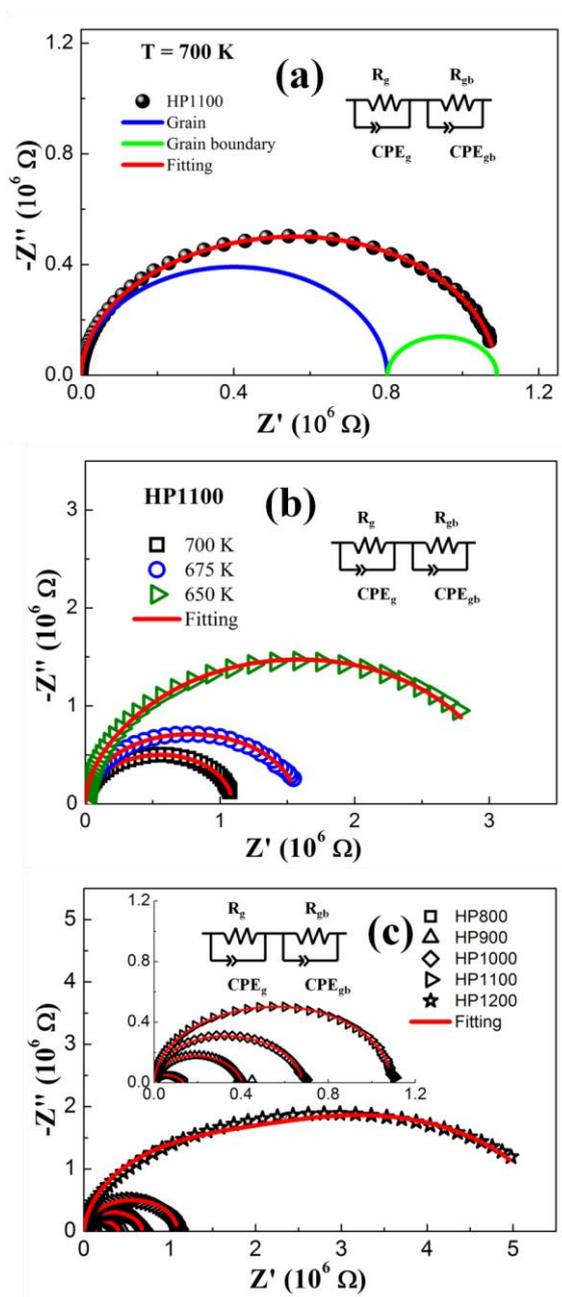

**Fig. 10. (a)** Impedance spectra of HP1100 measured at 700 K (black solid spheres) is fitted using equivalent circuit scheme using grain and grain boundary contributions, **(b)** Temperature evolution of impedance spectra of HP1100 measured at 650 K, 675 K, and 700 K fitted using equivalent circuit scheme, **(c)** Impedance spectra of different HP sintered ceramics measured at 700 K fitted using equivalent circuit scheme.

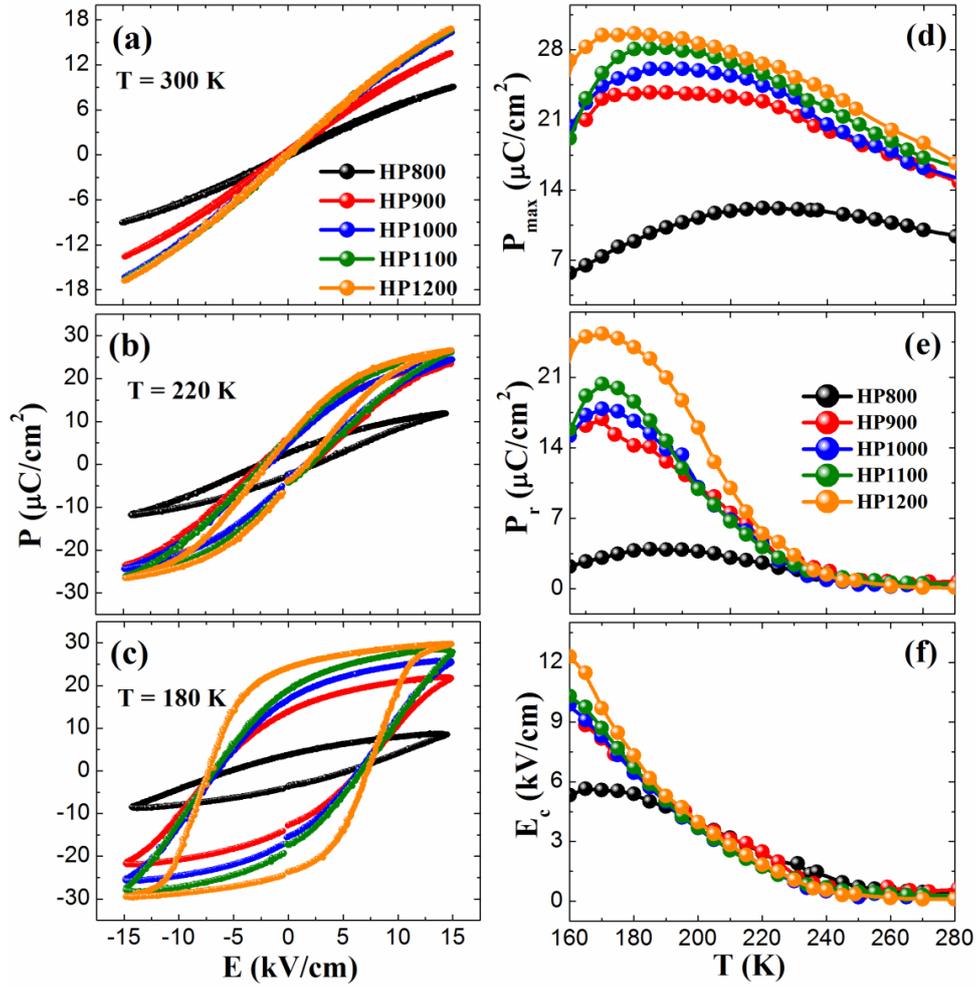

**Fig. 11. (a-c)** Comparison of PE hysteresis loop of HP sintered PMN ceramics (from 800 °C to 1200 °C) at E = 15 kV/cm and $f$ = 50 Hz **(a)** T = 300, **(b)** T = 220 K, and **(c)** T = 180 K, **(d-f)** Comparison of temperature dependent parameters **(d)** $P_{max}$, **(e)** $P_r$, and **(f)** $E_c$ of HP PMN ceramics.